\documentclass[12pt]{article}
\usepackage[dvips]{graphicx,rotating}
\usepackage{epsfig,rotating}
\usepackage{amsmath}
\begin{document}

\pagestyle{plain} 

\raggedbottom


{\bf \Large
  Resource paper: Nuclei with enhanced Schiff moments in practical elements for atomic and molecular  EDM measurements}

\vspace*{6mm}


This is a resource paper
concerning enhancement of
observable time-reversal breaking effects by nuclear structure,
aimed at AMO experimentalists.
It's intended to support
a white paper~\cite{Hutzler2022} by providing some
orientation on what can be said about some particular isotopes.
Any conclusions are qualitative, and the reader should consult and cite the
primary references and reviews rather than this arXiv alone.

\vspace*{9mm}
        {\em J.A. Behr, TRIUMF, 4004 Wesbrook Mall, Vancouver, BC V6T2A3 Canada}

        behr@triumf.ca

\section{Qualitative info on enhanced atomic and molecular EDM's from
  nuclear phenonmena}

Many nuclei have low-lying excited states with same spin as the ground state and opposite parity. A number of these have recently been suggested in the AMO literature as possibly exhibiting static or vibrational octupole deformation.
A goal is to identify practical candidates with enhanced
Schiff moments in particular elements where advanced AMO techniques could offer increased sensitivity to boosted atomic and molecular EDM's. Isotopes are sought in addition to $^{225}$Ra that are long-lived enough to use away from isotope production facilities~\cite{FlambaumDzuba2020,FlambaumFeldmeier2020} or are producible from generators like $^{223}$FrAg~\cite{DeMille2020,Klos2021}.
There is a similar need to choose quantifiably large Schiff moments in crystals with stable rare earth elements to optimize experiments searching for CP-violating couplings to long-wavelength axionlike dark matter~\cite{Arvanitaki2021}.
Some of the new parity doublets suggested are supported by
indirect experimental info, e.g. BE(3) strength measured or calculated in nearby even-even nuclei.
Unfortunately, other nuclear structure phenomena that would not enhance Schiff moments can
account for closely spaced parity doublets.
Whether or not the ground state and higher-spin states built on it look like one of a pair of static or vibrational octupole bands,
often there is limited information about the opposite-parity excited states.
Considerable effort over three decades has gone into identifying definitive
experimental nuclear signatures for static and vibrational octupole deformation for many nuclei~\cite{Ahmad1993,Butler1996,Butler2016,Butler2020a}.

Here we examine informally to what extent a number of these suggested
low-lying parity doublets would need more experimental nuclear structure
information and/or microscopic calculations to identify and quantify possible Schiff moment enhancement.
Whenever we list experimental evidence against static octupole deformation for a few
interesting AMO cases, we try to qualify to what extent collective octupole vibrations can have similar Schiff moment enhancement. Near degeneracy of a parity doublet will obviously enhance such effects in the energy denominator of first-order perturbation theory (an example is $^{19}$F, where accurate nuclear theory and an experimental limit on the Schiff moment exist~\cite{Ng2026}), but of course the governing matrix element must also be large. 

Note that Ref.~\cite{Hutzler2022} considers AMO methods  less element-specific, thus possibly applicable to any Z$\geq$86 nucleus
\footnote{This paper avoids the word `actinide,'
  used for `f-element' by AMO, more broadly by nuclear history}
 determined to have Schiff enhancements,
including quantum logic spectroscopy~\cite{Jayich2021},
frozen noble gas hosts~\cite{Singh2021},
and ultracold molecule clock states~\cite{Verma2020}.
The possible parity doublet in $^{229}$Pa has energy splitting 60$\pm$50 eV~\cite{Ahmad2015},
and experiments to resolve its status are considered~\cite{Singh2022}.

\subsection{TRV enhancements from nuclear structure: background}

{\bf A continuum between static octupole deformation and collective octupole vibrations.}
Schiff moment-induced atomic and molecular EDM's are thought to be similar in systems with static octupole deformation and in systems with `soft' (i.e. virtual) excitation of octupole vibrations that are more common, though those octupole vibrations must be collective to helpfully enhance the Schiff moment. Engel, Friar, and Hayes (EFH)~\cite{Engel2000} discuss to what extent these extremes are different aspects of formally similar phenomena (see other reviews like Ref.~\cite{Butler2016} and Ref.~\cite{Engel2025} for related discussions of octupole vibrations and static deformation), and showed that in macroscopic models a Schiff moment enhancement from collective octupole vibrations also scales with the inverse of the parity doublet's energy splitting, with the static deformation $\langle \beta_3 \rangle^2$ replaced by the virtual deformation average $\langle {\bar \beta_3}^2 \rangle$.
EFH point out additional physical effects that must be calculated for the vibrators, which could enhance the Schiff moment further.
Flambaum and Zelenvinsky apply a similar analytical calculation of collective octupole vibrations to cases of interest $^{223}$Rn and $^{223,225}$Ra, finding similar Schiff enhancement compared to previous calculations assuming static octupole deformation~\cite{Flambaum2003}.
Auerbach {\em et al.}~\cite{Auerbach2006} with microscopic QRPA find similarly large Schiff moment enhancement from collective vibrations in the limit when their energies are small: they find in Z$\geq$86 nuclei (with N and Z separated from the region of known static octupole deformation) that octupole vibrational configurations are only small components of the complex nuclear wavefunctions, suppressing the Schiff moment enhancement. Note that configurations are assumed purely octupole in some analytical approximation formulae, though Ref.~\cite{Spevak1997} calculates such a suppression factor.
This author is unaware of any calculations that consider interference between
static and collective vibrational effects, which presumably would be included naturally in future microscopic calculations~\cite{Engel2022}.

{\bf Nuclear magnetic quadrupole moments}
The possible enhancements of Schiff moments between two and five
orders of magnitude from static octupole deformation have produced 
much attention in designing AMO experiments.
Additionally,
atomic and molecular EDM's can be enhanced by other nuclear effects by
about an order of magnitude, potentially leading to exciting discovery
potential if the AMO experiment has sufficient sensitivity.
Contributions of the more ubiquitous nuclear magnetic quadrupole moment (MQM)
to atomic and molecular EDM's are free from atomic shielding, so
have been reliably calculated~\cite{Haxton1983,Sushkov1984}.
MQM's provide order-of-magnitude enhancement 
for any nucleus with quadrupole deformation and nuclear spin $\geq$ 1,
and if the atomic or molecular state has nonzero angular momentum.
Experimental difficulties from the nonzero electron spin needed 
can be mitigated by choosing systems with small magnetic g-factor, allowing in
principle long coherence times, and its known existence in many
stable nuclei  is driving dedicated experimental
efforts
~\cite{Hutzler2020,Hutzler2022}.


{\bf TRV matrix elements}
That all atomic EDM enhancements from nuclear structure also require
difficult-to-calculate matrix elements of parameterized TRV hadronic
interactions was made apparent in Ref.~\cite{Dobaczewski2018}, with possible microscopic theoretical improvements of the spatial wavefunctions necessary highlighted in Ref.~\cite{Engel2022}. 
Some analytical formulae include an estimate of one TRV matrix element from a one-body effective potential~\cite{Spevak1997,Sushkov1984}. There are no known helpful experimental benchmarks for the most important operator $\sigma \cdot p$ in heavy nuclei.

\subsection{Observables}
Consult the reviews
for more complete information on observables.

{\bf Charge radii changes with neutrons} Adding a neutron to a nucleus increases its charge radius. For a given element,
adding a neutron to an odd-N nucleus usually increases the charge radius
by less than adding a neutron to an even-N nucleus-- intuitively, this is
reasonable when the extra neutron simply pairs in spin with the odd valence
neutron in the same orbital, rather than populating another orbital or driving a change in collective properties. Octupole deformation is a possible explanation for unusual isotopes with inverted differences in neutron-induced change of charge radii (``IDNCR'') in Rn,Fr,Ra~\cite{Otten1989,Ahmad1988,Leander1982}, with measurements further delineating IDNCR at higher neutron number~\cite{Budincevic2014,Lynch2018} and in Ac~\cite{Verstraelen2019}. Other explanations than octupole phenomena in lower-Z nuclei are considered in Ref.~\cite{Perera2021}, which also mentions the possibility that collective vibrational effects might similarly produce IDNCR. For some isotopes, IDNCR is the only experimental information, yet thought it can be consistent with
static octupole deformation or collective octupole vibrations,
unfortunately can't be definitive.

{\bf BE(3)'s, BE(1)'s} A possible smoking gun is electric octupole transition strength B(E3) between the opposite-parity bands, but AMO people 
should recall the long-wavelength expansion works really well in nuclei
(r/$\lambda$ $\sim$ 1 MeV/197 MeV-fm $\sim$ 0.5\%). So if an E1 is allowed, the E3's
start smaller by over 8 orders of magnitude and are usually considered unmeasurable.
Thus measurements of E3 strength between the ground
$0^+$ state and excited 3$^-$ state in nearby even-even nuclei are
possible and useful~\cite{Gaffney2013,Butler2020}.

Strong B(E1)'s are expected between the opposite-parity bands 
of static octupole deformation,
while weak B(E1)'s are often considered to be consistent with collective octupole vibrations.
The enhancements can be one or two orders of magnitude larger than the typical
10$^{-3}$ to 10$^{-4}$ Weiskopf units (WU) of low-lying E1 transitions~\cite{Ahmad1993}. Given that a typical giant dipole resonance that has collected most of the E1 strength has 10 WU, other physics like single-particle transitions can produce large enough E1 amplitudes to interfere constructively or destructively with octupole-driven effects.
Since there can be other reasons for either the enhancements or the lack thereof, B(E1)'s are not considered definitive evidence of octupole physics unless there are thorough theory considerations.
Flambaum and Mansour have used the ``lightning rod'' macroscopic concept of charge collecting at sharper curvature that drives E1 enhancement to estimate the resulting Schiff moment~\cite{Mansour2025}, and some somewhat challenging results will be considered below.

{\bf Odd-odd nuclei}
Note that
nuclei with both odd N and odd Z can have octupole deformation~\cite{Sheline1988b}, possibly doubling the number of cases with Schiff enhancement.  However, the nuclear structure is so much more complicated
than odd-even nuclei that it's much less likely the Schiff enhancement could be definitively determined, either experimentally or theoretically.


\section{Examples}
Here are considerations of isotopes of AMO interest.

\subsection{Rare earths}
A number of rare earth elements have been laser-cooled and trapped, and
several isotopes of such elements with low-lying parity doublets are being
suggested for static octupole deformation~\cite{FlambaumDzuba2020,FlambaumFeldmeier2020}.

{\bf $^{153}$Eu has been examined thoroughly.}
$^{152,154}$Eu are odd Z- odd N nuclei with half-lives of 13 and 8 years that have structure and E1 strengths suggesting octupole deformation, while they and stable $^{153}$Eu exhibit IDNCR consistent with octupole deformation in the ground states~\cite{Sheline1989}.
Such evidence inspired detailed experimental examination of excited states in $^{153}$Eu, and
the g-factors of the identified excited opposite-parity band are known to be about
3 times larger than the ground state band~\cite{Smith1998,Pearson1994}.
Now, it's natural for perturbations
such as Coriolis mixing
to change properties of partner bands differently-- nearby configurations
tend to have the natural parity of the ground-state--
thus certain properties don't have to be identical.
Nevertheless, such different g-factors discourage an interpretation of this
doublet in $^{153}$Eu as exhibiting static octupole deformation, while relatively large E1's between the bands are hard to interpret in terms of octupole vibrations~\cite{Smith1998,Pearson1994}. So $^{153}$Eu remains ambiguous, and needs
further modelling and calculation to determine whether its identified parity
doublet has octupole phenomena.

{\bfseries Theory and experiment updates for $^{153}$Eu:} Sushkov, after pointing out the above differing magnetic moments of the $^{153}$Eu 5/2$^+$ ground state and low-lying 5/2$^-$ state are inconsistent with the static octupole model, calculates the admixture of octupole vibration to deformed Nilsson states, relating the Schiff moment to experimental data on magnetic moments and E1 and E3 transition amplitudes in odd-even nuclei and adjacent even-even nuclei. $^{153}$Eu is one of the few nuclei with all data needed, and Ref.~\cite{Sushkov2024} finds 30x enhancement of the resulting Schiff moment of $^{153}$Eu over spherical nuclei and estimates model uncertainty, finding a similar result for $^{237}$Np. A recent calculation with density functional-based theory has some disagreement on the Schiff moment of $^{153}$Eu, mostly from finite-ranged interactions~\cite{Engel2024}. The macroscopic relation between B(E1) and Schiff moment in Ref.~\cite{Mansour2025} arrives at a similar answer.

A newly published experiment of a null Schiff moment for $^{153}$Eu with complementary sensitivity~\cite{Vutha2025} will hopefully drive more definitive theory calculations with well-defined uncertainty.  

{\bf Other stable rare earths}
Several other stable rare earth isotopes have low-lying parity doublets
($^{161,163}$Dy, $^{165}$Er, $^{155}$Gd, $^{153}$Sm)~\cite{FlambaumDzuba2020,FlambaumFeldmeier2020},
while nearby even-even nuclei with measured 3$^-$ states and BE(3) strength support the possibility of collective vibrational octupole configurations in such nuclei as well as in $^{153}$Eu~\cite{Kibedi2002}.
A recent 
global calculation of 
particle-bound even Z - even N nuclei
predicts static octupole deformation in the ground states of
stable $^{146,148,150}$Nd and $^{150}$Sm~\cite{Cao2020},
while a region in Z and N of radioactive rare earth nuclei with much higher
N than
the line of stability has been identified to have static octupole
deformation, both experimentally and theoretically~\cite{Cao2020,Butler2016,Butler2020}.
These stable rare earth nuclei proposed for Schiff moment enhancements need further experimental and/or microscopic theoretical supporting evidence for static or collective vibrational octupole deformation in their ground and excited states.

\subsection{ Z$\geq$86}

\begin{figure}[hbtp]
 	\centering
 	\includegraphics[width=0.8\textwidth]{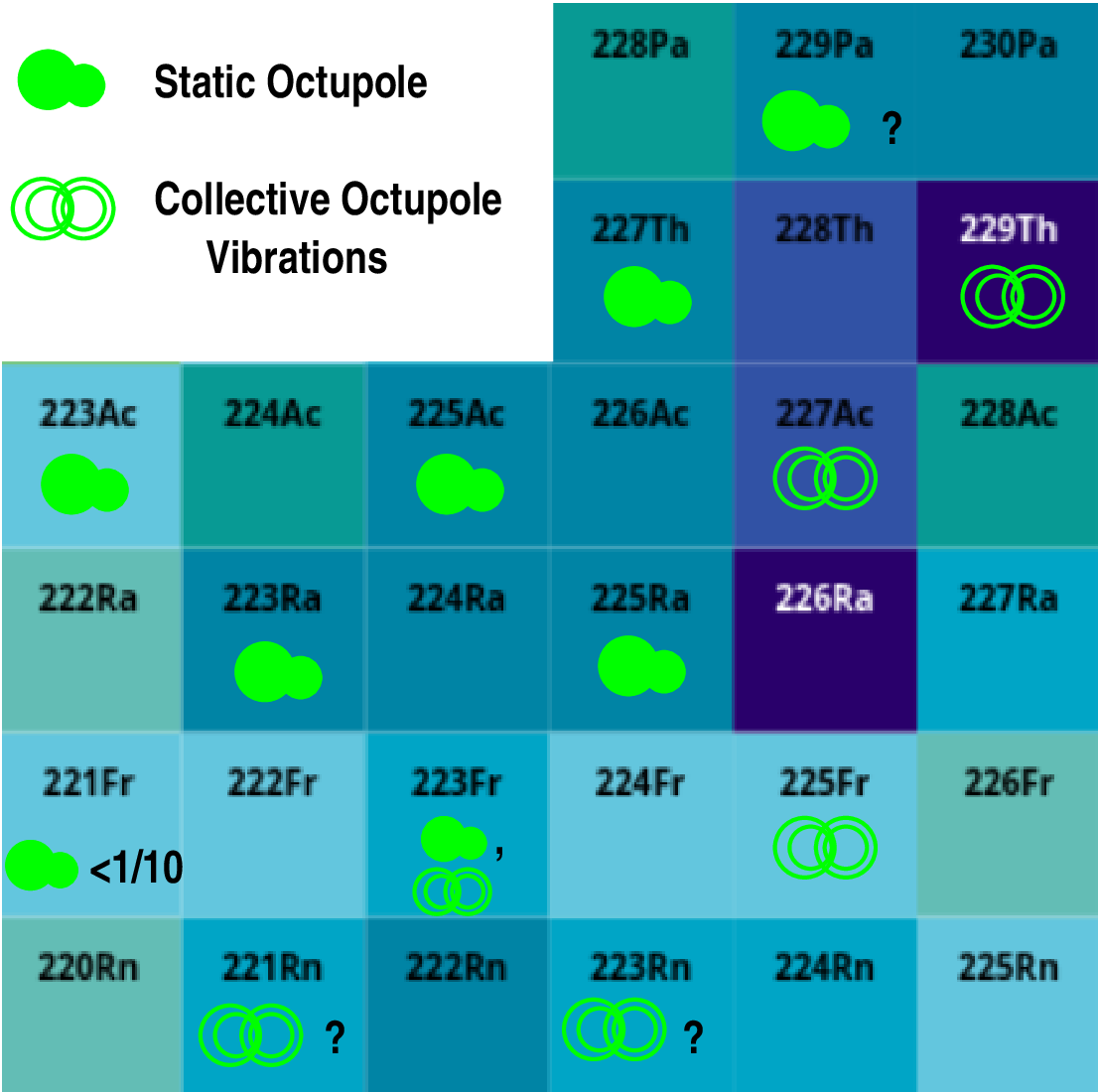}
 	\caption{Overly simplistic summary of Z$\geq$86 isotopes of possible AMO interest that have experimental evidence suggesting extremes in the continuum of static octupole deformation vs. collective octupole vibrations, either of which can enhance TRV Schiff moments.
 	Only the dominant configuration is indicated-- see text and references for any details. Darker blues have longer half-lives (see nndc.bnl.gov).}
 	\label{fig:OctIsotopes}
\end{figure}

AMO experiments have been considered for 
nuclei with supporting experimental evidence for static octupole
deformation and/or collective octupole vibrations for ground and excited
states of a parity doublet,
including $^{225}$Ra, $^{223}$Fr, $^{225,227}$Ac, and $^{227,229}$Th (see Fig.~\ref{fig:OctIsotopes}).

{\bf Radium:} The solid experimental evidence supporting static octupole
deformation in I=1/2 $^{225}$Ra and I=3/2 $^{223}$Ra is
thoroughly documented in review articles~\cite{Butler2016,Butler1996}.

{\bf Francium}: $^{223}$Fr is an illustrative case where experiment and theory suggest a large enhancement of the Schiff moment, even though the nuclear wavefunctions are thought to contain significant admixtures of static octupole deformation and collective octupole vibrations.
IDNCR exists for $^{223,225}$Fr~\cite{Otten1989,Coc1987} (it barely exists for
$^{221}$Fr, a conclusion that does not change with a slightly revised optical isotope shift~\cite{Lu1997}).
Concerning excited states, it's instructive to note that
$^{221}$Fr is a rather complex ``transition'' nucleus; e.g., its
possible static octupole deformed excited parity-doublet band
actually has spins out of order in energy.
This is understood as a case in neutron number N between $^{219}$Fr,
which has no special deformation,
and $^{223}$Fr, which has ``fairly strong'' evidence~\cite{Sheline1995} from its structure and feeding from alpha- and beta- decay for a static octupole deformation parity doublet, albeit with a relatively large parity-doublet splitting of 161 keV.
The experimental parity doublet of $^{223}$Fr has been modelled including admixtures with static octupole and collective octupole vibrations~\cite{Sheline1995}; however,
E1 transitions between the $^{223}$Fr bands were later found to be relatively weak, consistent with $^{223}$Fr having collective octupole vibrations~\cite{Kurcewicz1992,Kurcewicz1992erratum,Burke1997}.
Ref.~\cite{Spevak1997}, which considers theoretical admixtures of other configurations in all its calculations, predicts the static octupole enhancement of EDM's for $^{221}$Fr
to be an order of magnitude less than $^{223}$Fr and $^{225}$Ra.
$^{225}$Fr has weak B(E1)'s inconsistent with static octupole deformation but consistent with collective octupole vibrations~\cite{Burke1997}. (Note Ref.~\cite{Burke1997} has a thorough
pedagogical discussion of the B(E1) phenomenology, including an updated figure of the B(E1)'s in Ref.~\cite{Ahmad1993}.)

Because of ambiguities in interpreting B(E1)'s, the ``lightning rod'' estimate of Schiff moments of Ref.~\cite{Mansour2025} does not provide better definition for $^{221}$Fr or $^{223}$Fr. The update in the relevant $^{223}$Fr B(E1) value above, which produces a tiny Schiff moment enhancement, is acknowledged in the arXiv version of Ref.~\cite{Mansour2025}. The relatively strong $^{221}$Fr B(E1) used in Ref.~\cite{Mansour2025} turns out to be from an octupole parity doublet unrelated to the ground state [A. Mansour, private comm.] and the $^{221}$Fr B(E1) derived from a B(E2) in Ref.~\cite{ButlerE1} uses a macroscopic relation that implicitly assumes a wavefunction dominated by a large Schiff moment enhancement.

{\bf Actinium}:
I$^\pi$=3/2$^-$ $^{227}$Ac has a low-lying 3/2$^+$ excited state, reviewed in~\cite{Ahmad1993}. There is considerable evidence consistent with these being
the lowest-lying states of bands with static octupole
deformation, including measurements of g-factors. But there is also 
experimental evidence against static octupole deformation for this parity doublet, including
measured BE(1)'s an order of magnitude smaller than other cases, and
transfer reactions populating excited states that can be explained by reflection-symmetric potentials~\cite{Ahmad1993}. $^{227}$Ac could be a case with
large collective vibrational octupole enhancement, and EDM experiments involving actinium molecules~\cite{FlambaumDzuba2020} could also be done in
$^{225}$Ac, which  has strong supporting evidence for static octupole deformation~\cite{Ahmad1993} and a 10-day half-life producible by a $^{229}$Th generator.

{\bf Thorium}: Some EDM experimental techniques work best for nuclear spin I=1/2. In addition to the well-known I=1/2 $^{225}$Ra, $^{227}$Th likely has I=1/2~\cite{Kovalik2021}.
There is strong experimental and theoretical support for
static octupole deformation with an identified parity double band for an I=1/2 ground state, 
and a large Schiff moment enhancement~\cite{Butler1996,Butler2016}.

An explicit band of excited states is identified in $^{229}$Th as a vibrational
octupole partner of a ground state band, with energy splitting 146 keV~\cite{Ruchowska2006,Gulda2002}, and thus from the general expectations of Ref.~\cite{Engel2000} a full microscopic calculation could show a large Schiff moment for $^{229}$Th.
It is in this sense that
recent BE(1) measurements in $^{228}$Th
supporting vibrational collective octupole effects in $^{229}$Th are considered to support
it as an enhanced atomic or molecular EDM case~\cite{Chisti2020},
even though
detailed consideration of the excited states of $^{229}$Th has classified
it as a ‘transition’ isotope with larger N than nuclei with demonstrated static octupole deformation~\cite{Leander1984,Butler1996}.

{\bf Radon}:
IDNCR exists for
$^{221,223}$Rn~\cite{Otten1989},
but there is only one excited state energy known for $^{221}$Rn and no excited
state info for $^{223}$Rn.
Favorable static octupole 
arguments from early calculations~\cite{Spevak1997}
have been more recently weakened by BE(3) measurements in nearby
even-even nuclei~\cite{Butler2016},
which instead suggest collective vibrational octupole effects are possible~\cite{Butler2019}.
To quantify the Schiff enhancements would either way need detailed
excited-state nuclear structure measurements, though this has remained elusive in several experimental attempts. 

{\bf Summary}
All nuclei with reasonably definitive supporting experimental evidence for static octupole deformation are radioactive,
with half-lives on the order of weeks or shorter. 
Longer-lived nuclei like $^{229}$Th and $^{227}$Ac, along with stable rare earth isotopes, show evidence for collective octupole vibrations that could also produce quantifiably large Schiff enhancements, if microscopic calculations with sufficiently small uncertainty can be developed.
Any prospective parity doublet needs support from experimental observables to define what fraction of the ground state and excited states have static or collective vibrational octupole configurations. Microscopic theory improvements of this aspect, as well as to quantify matrix elements of the several TRV isospin- and range-dependent interactions~\cite{Engel2022}, would enable a move from the current exciting discovery phase to defining multiple complementary experiments to extract different microscopic sources of TRV.

{\bf Acknowledgements} P. Butler, J. Engel, N. Hutzler, A. Jayich, D. DeMille,
A. Vutha, W. Nazarewicz, A. Madden, G. Neyens, and G. Hackman all have had helpful comments and/or orientation. The author notes that two nuclear theorists calculating Schiff moments share surnames with AMO experimentalists in this field-- the AMO community should be inspired to encourage any further relatives to help. This author is, of course, responsible for any misinterpretations of the literature.

\end{document}